# Transitional layer at the edge of a false vacuum in a cavitation model of the Big Bang

Mikhail Pekker, Mikhail N. Shneider

*Department of Mechanical and Aerospace Engineering, Princeton University, Princeton, NJ USA*

m.n.shneider@gmail.com

## Abstract

This paper considers the structure and physical processes in the transition region at the border between the regions of false and physical vacuums in a cavitation model of the inflationary stage of the Big Bang. It is shown that in the process of the formation of physical vacuum bubbles in a false vacuum, conditions for the formation of a narrow layer filled with matter arise in the transition region, which is the precursor to bridges in the observed large-scale cellular structure of the universe.

## I. Introduction

In [1], we proposed a cavitation model of the inflationary stage of the Big Bang. In this model, following the work of Guth [2], the effect of the cosmological Λ-term is considered as a source of negative pressure acting in the region of a false vacuum, in which, as a result of the tunneling phase transition, bubbles of the physical vacuum appear, in which the cosmological constant is equal to zero. This process is similar to the tunneling regime of transition to cavitation in cryogenic liquid helium in the region of negative pressure [3]. The cavitation model of the inflationary stage [1] makes it possible to explain the homogeneity and large-scale isotropy of the universe without generating unrelated space-time universes and to explain the large-scale cellular structure of the universe.

In this work, it is shown that in the transition layer at the border of the false vacuum and arising bubbles of the physical vacuum, the formation of narrow layers of matter is possible, which are potential precursors to bridges in the observed large-scale cellular structure of the universe.

## II. The boundary of the real vacuum bubble

Consider the space-time metric in the frame of reference associated with the center of a spherically spherical physical (true) vacuum bubble surrounded by a false vacuum for a time much less than the characteristic time of the expansion of the universe $t \ll t_i$, where $t_i$ is the characteristic time of the expansion of the universe in standard inflationary models [2,4-12]. As in [1], the transition region at the border of the false and physical vacuums will be considered in the De Sitter metric. In accordance with [13], the expression for the interval *ds* in the centrally symmetric case in terms of the variables $t$, $r$, and $\theta$, $\phi$ has the form

$$ds^2 = e^{\nu} c^2 dt^2 - e^{\lambda} dr^2 - r^2 (d\theta^2 + sin^2\theta d\phi^2). \tag{1}$$

In this case, Einstein's equations are converted to form [13]

$$\frac{8\pi G}{c^4}T_0^0 = \frac{8\pi G}{c^4}\tilde{T}_0^0 - \Lambda = -e^{-\lambda}\left(\frac{1}{r^2} - \frac{\lambda'}{r}\right) + \frac{1}{r^2} \qquad (2)$$

$$\frac{8\pi G}{c^4}T_1^1 = \frac{8\pi G}{c^4}\tilde{T}_1^1 - \Lambda = -e^{-\lambda}\left(\frac{\nu'}{r} + \frac{1}{r^2}\right) + \frac{1}{r^2} \qquad (3)$$

$$\frac{8\pi G}{c^4}T_2^2 = \frac{8\pi G}{c^4}\tilde{T}_2^2 - \Lambda = \frac{8\pi G}{c^4}\tilde{T}_3^3 - \Lambda = -\frac{1}{2}e^{-\lambda}\left(\nu'' + \frac{\nu'^2}{2} + \frac{\nu' - \lambda'}{r} - \frac{\nu'\lambda'}{2}\right) + \frac{1}{2}e^{-\nu}\left(\ddot{\lambda} + \frac{\dot{\lambda}^2}{2} - \frac{\dot{\lambda}\dot{\nu}}{2}\right) \qquad (4)$$

$$\frac{8\pi G}{c^4}T_0^1 = \frac{8\pi G}{c^4}T_1^0 = -e^{-\lambda}\frac{\dot{\lambda}}{r}. \qquad (5)$$

B (2)–(5) $c$ is the speed of light, the dots above $\lambda$ and $\nu$ are the derivatives with respect to time, and the primes are the derivative with respect to the coordinate $r$. In (2)–(4), we have identified in the energy-momentum tensor $T_k^i$ the cosmological term $\Lambda$, introduced by Einstein in the equations of the General Theory of Relativity (see, for example, [13]).

We will look for a stationary solution (2)–(5) in the regions of false ($r \geq R_b$) and physical ($r < R_b$) vacuums, where $R_b$ is the radius of the formed bubble of the physical vacuum (Fig. 1). We will assume that $\tilde{T}_k^i = 0$ and $\Lambda = \Lambda_\infty = \text{const} \neq 0$ far from the boundary in the false vacuum region ($r \gg R_b$), and $T_k^i = 0$ in the physical vacuum region.

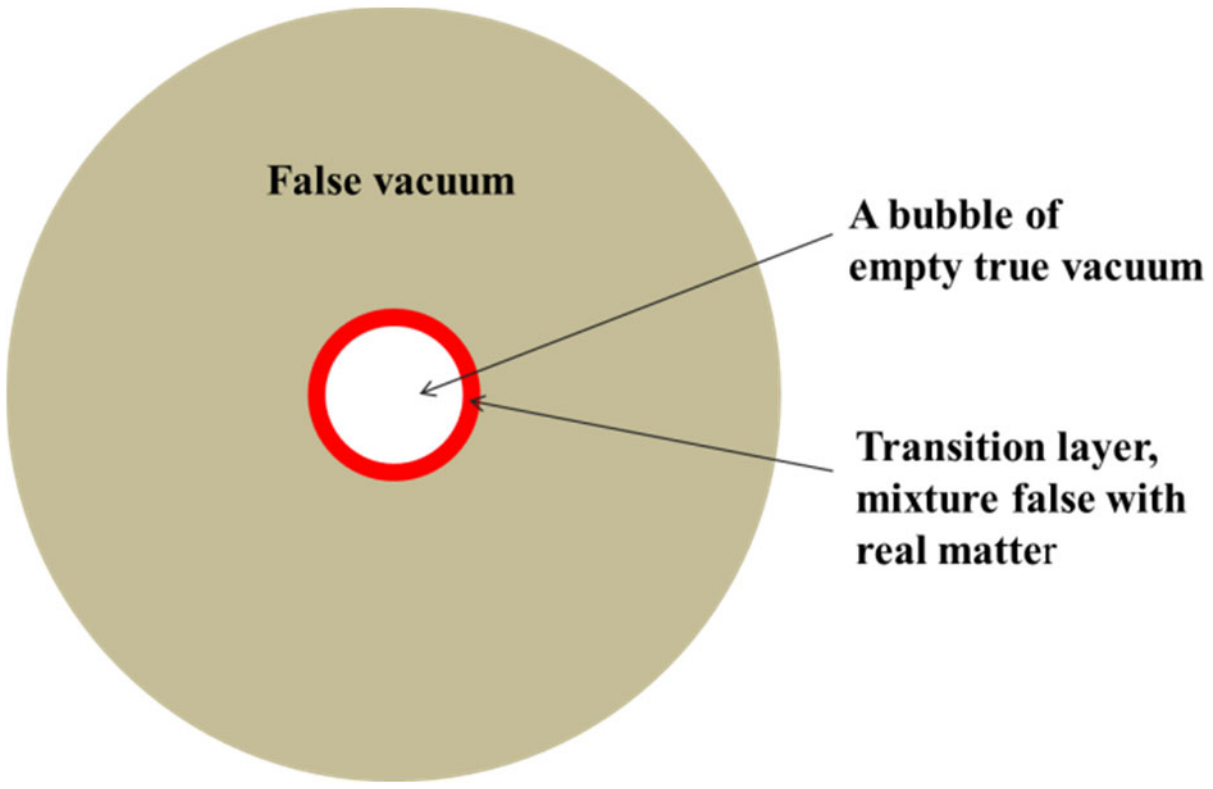

Fig.1. A bubble of true vacuum inside the false vacuum.

Let us show that it is impossible to match solutions on the boundary of physical and fake vacuums (Fig. 1) without taking into account $\tilde{T}_k^i$ in (2)–(4). Indeed, in the stationary case, equations (2)–(4) have the form

$$-\Lambda = -e^{-\lambda}\left(\frac{1}{r^2} - \frac{\lambda'}{r}\right) + \frac{1}{r^2} \qquad (6)$$

$$-\Lambda = -e^{-\lambda}\left(\frac{\nu'}{r} + \frac{1}{r^2}\right) + \frac{1}{r^2} \qquad (7)$$

$$-\Lambda = -\frac{1}{2}e^{-\lambda}\left(\nu'' + \frac{\nu'^2}{2} + \frac{\nu' - \lambda'}{r} - \frac{\nu'\lambda'}{2}\right). \qquad (8)$$

Subtracting (7) from (6), we obtain, $\lambda' + \nu' = 0$. Substituting $\lambda' = -\nu'$ in (7) and (8), we have

$$-\Lambda = -e^{\nu}\left(\frac{\nu'}{r} + \frac{1}{r^2}\right) + \frac{1}{r^2} \qquad (9)$$

$$-\Lambda = -\frac{1}{2}e^{\nu}\left(\nu'' + \nu'^2 + 2\frac{\nu'}{r}\right). \tag{10}$$

Multiplying (9) by $r^2$ and differentiating along the radius, we get

$$-\left(\Lambda + \frac{1}{2}\Lambda' r\right) = -\frac{1}{2}e^{\nu}\left(\nu'' + \nu'^2 + 2\frac{\nu'}{r}\right). \tag{11}$$

When obtaining (11), we took into account that in the transition region $\Lambda = \Lambda(r)$. Comparing the left-hand sides of (11) and (10) (the right-hand sides are the same), we get that $\Lambda' \equiv 0$. In other words, it is impossible to describe the region of transition of a real vacuum to a false one by one function $\Lambda$. The formation of a bubble of a physical vacuum always occurs with the formation of matter, characterized by the energy-momentum tensor $\tilde{T}_k^i$.

As in [1], we put $\nu = -\lambda$. In this case, the system of equations (2)–(5) is reduced to the following equations:

$$\frac{8\pi G}{c^4}\tilde{T}_0^0 - \Lambda = -e^{-\lambda}\left(\frac{1}{r^2} - \frac{\lambda'}{r}\right) + \frac{1}{r^2} \tag{12}$$

$$\frac{8\pi G}{c^4}\tilde{T}_2^2 - \Lambda = -\frac{1}{2}e^{-\lambda}\left(-\lambda'' + \lambda'^2 - 2\frac{\lambda'}{r}\right). \tag{13}$$

Recall that $\tilde{T}_3^3 = \tilde{T}_2^2$, $\tilde{T}_1^1 = \tilde{T}_0^0$ and that all other components of $\tilde{T}_k^i$ are equal to zero. We rewrite (12) and (13) as follows:

$$\frac{d}{dr}\left(re^{-\lambda}\right) = 1 - r^2\left(\frac{8\pi G}{c^4}\tilde{T}_0^0 - \Lambda\right) \tag{14}$$

$$\frac{8\pi G}{c^4}\tilde{T}_2^2 = -\frac{1}{2r}\frac{d^2}{dr^2}\left(re^{-\lambda}\right) + \Lambda = \frac{8\pi G}{c^4}\tilde{T}_0^0 + \frac{r}{2}\left(\frac{8\pi G}{c^4}\frac{d\tilde{T}_0^0}{dr} - \frac{d\Lambda}{dr}\right). \tag{15}$$

From (14), it follows that

$$e^{-\lambda} = 1 - \frac{1}{r}\int_0^r r^2\left(\frac{8\pi G}{c^4}\tilde{T}_0^0 - \Lambda\right)dr. \tag{16}$$

To match Eq. (14) in the region of the false vacuum $r > R_b$ with the solution in the region of the physical $r < R_b$, we will assume that the size of the transition region from the false vacuum to the physical $\delta$ is much less than $R_b$. In this case, one can put

$$\Lambda = \frac{1}{2}\Lambda_\infty\left(1 + erf\left(\frac{r-R_b}{\delta}\right)\right) = \Lambda_\infty\frac{1}{2}\left(1 + \frac{2}{\sqrt{\pi}}\int_0^{\frac{r-R_b}{\delta}} e^{-y^2}dy\right) \tag{17}$$

$$\tilde{T}_0^0 = \frac{1}{4}\eta\left(1 - erf^2\left(\frac{r-R_b}{\delta}\right)\right) = \eta\frac{1}{4}\left(1 - \left(\frac{2}{\sqrt{\pi}}\int_0^{\frac{r-R_b}{\delta}} e^{-y^2}dy\right)^2\right), \tag{18}$$

where $erf(x) = \frac{2}{\sqrt{\pi}}\int_0^x e^{-y^2}dy$ is the error function, and $\eta$ is an independent parameter of the problem. We have taken into account that in the false vacuum region, far from the interface, $\tilde{T}_0^0 = 0$, and the curvature is determined exclusively by the cosmological constant $\Lambda$.

For the convenience of the analysis, let us move on to dimensionless variables, as follows:

$$\Lambda_* = \frac{\Lambda}{\Lambda_\infty}, T_{*0}^0 = \frac{8\pi G}{c^4}\frac{\tilde{T}_0^0}{\Lambda_\infty}, T_{*2}^2 = \frac{8\pi G}{c^4}\frac{\tilde{T}_2^2}{\Lambda_\infty}\ x = \frac{r}{r_0}, r_0 = \frac{1}{\sqrt{\Lambda_\infty}}, \eta_* = \frac{8\pi G}{c^4}\frac{\eta}{\Lambda_\infty},\ \ R_* = \frac{R_b}{r_0},\ \delta_* = \frac{\delta}{r_0}. \tag{19}$$

Substituting (17) and (18) into (15) and (16), we obtain in dimensionless variables (19)

$$e^{-\lambda} = 1 - \frac{1}{x}\int_0^x x^2\left(\frac{\eta_*}{4}\left(1 - erf^2\left(\frac{x-R_*}{\delta_*}\right)\right) - \frac{1}{2}\left(1 + erf\left(\frac{x-R_*}{\delta_*}\right)\right)\right)dx \tag{20}$$

$$T_{*2}^2 = T_{*0}^0 + \frac{x}{2}\left(\frac{dT_{*0}^0}{dx} - \frac{d\Lambda_*}{dx}\right). \tag{21}$$

Figure 2 shows the dependences of $\Lambda_*$, $T_{*0}^0$и $T_{*2}^2$ on $x$.

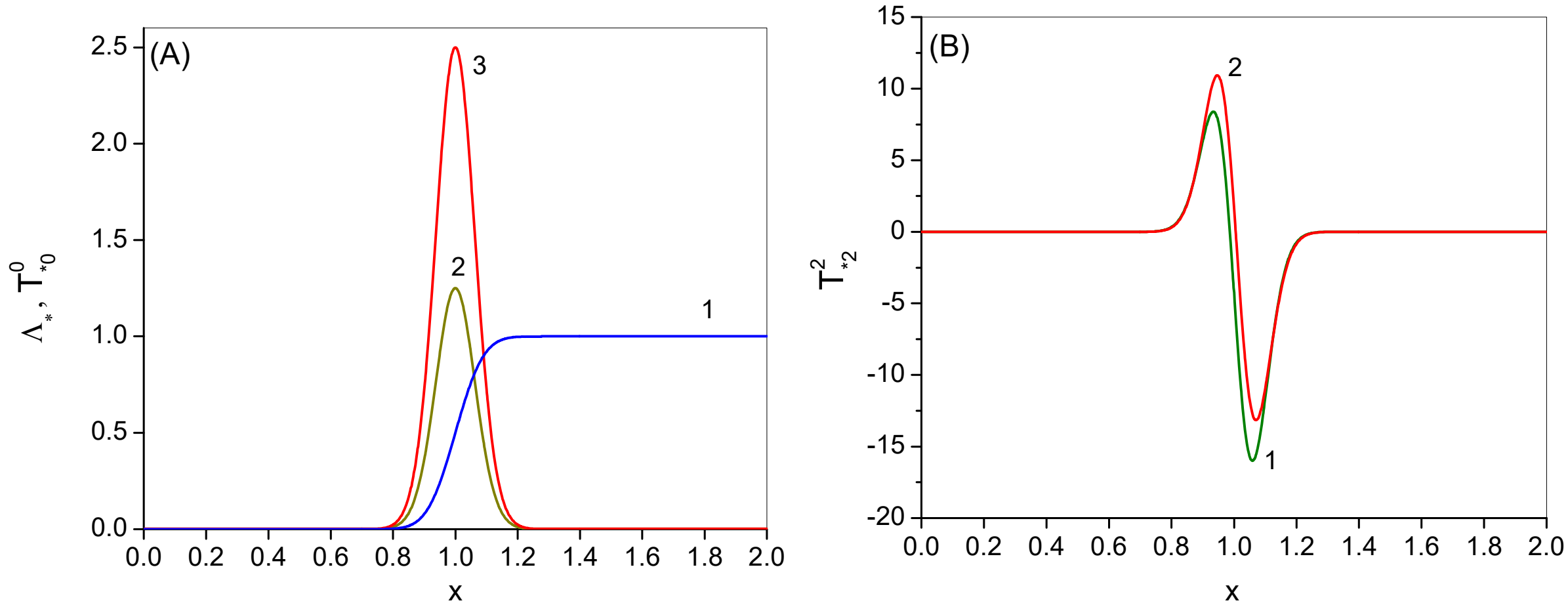

Fig. 2. In Fig. (A), line 1 corresponds to $\Lambda_*$, line 2 to $T_{*0}^0$ at $\eta_* = 5$, and line 3 to $T_{*0}^0$ at $\eta_* = 10$. In Fig. (B), line 1 corresponds to $T_{*2}^2$ at $\eta_* = 5$, and line 2 to $T_{*2}^2$ at $\eta_* = 10$; $\delta_* = 0.1$, $R_* = 1$.

Consider the forces acting on particles in the region of false and real vacuums. Considering the gravitational field at the discontinuity to be independent of time, the force acting on a particle of mass $m$ in the centrally symmetric case, according to [13, chapter 88, problem 1, formula (3)] for interval (1), is

$$f = -Mc^2\frac{\partial}{\partial r}\left(ln\sqrt{g_{00}}\right) = -\frac{Mc^2}{R_b}\cdot\frac{\frac{1}{x^2}\int_0^x y^2(T_{*0}^0 - \Lambda_*)dy - x(T_{*0}^0 - \Lambda_*)}{2\left(1 - \frac{1}{x}\int_0^x y^2(T_{*0}^0 - \Lambda_*)dy\right)}, \tag{22}$$

where $M = m/(1 - v^2/c^2)$ is the relativistic mass of the particle ($m$ is the rest mass, $v$ is the velocity). In (22), we took into account that the component of the metric tensor $g_{00} = e^{-\lambda}$ (see (1)).

The potential energy of the particles corresponding to the force $f$ is equal to

$$U = -\int_0^r f dr = Mc^2 ln\sqrt{g_{00}}. \tag{23}$$

At $\frac{x-R_*}{\delta_*} \gg 1$, the contribution of $T^0_{*0}$ to (20) can be neglected. Using the asymptotic expression for the metric tensor component at $x \gg R_*$, $g_{00,a} = e^{-\lambda_{as}} = 1 + \frac{R_*^2}{3}x^2$, we write out the asymptotic values for $f$ and $U$ obtained in [13]:

$$f_a = -\frac{Mc^2}{R_b}\frac{R_*^2 x}{(3+R_*^2 x)}, \quad U_a = \frac{1}{2}Mc^2 ln\left(1 + \frac{R_*^2}{3}x^2\right). \tag{24}$$

Figure 3 shows the dependences of the reduced values $f_* = f/(Mc^2/R_b)$ and $U_* = U/(Mc^2)$ on $x$.

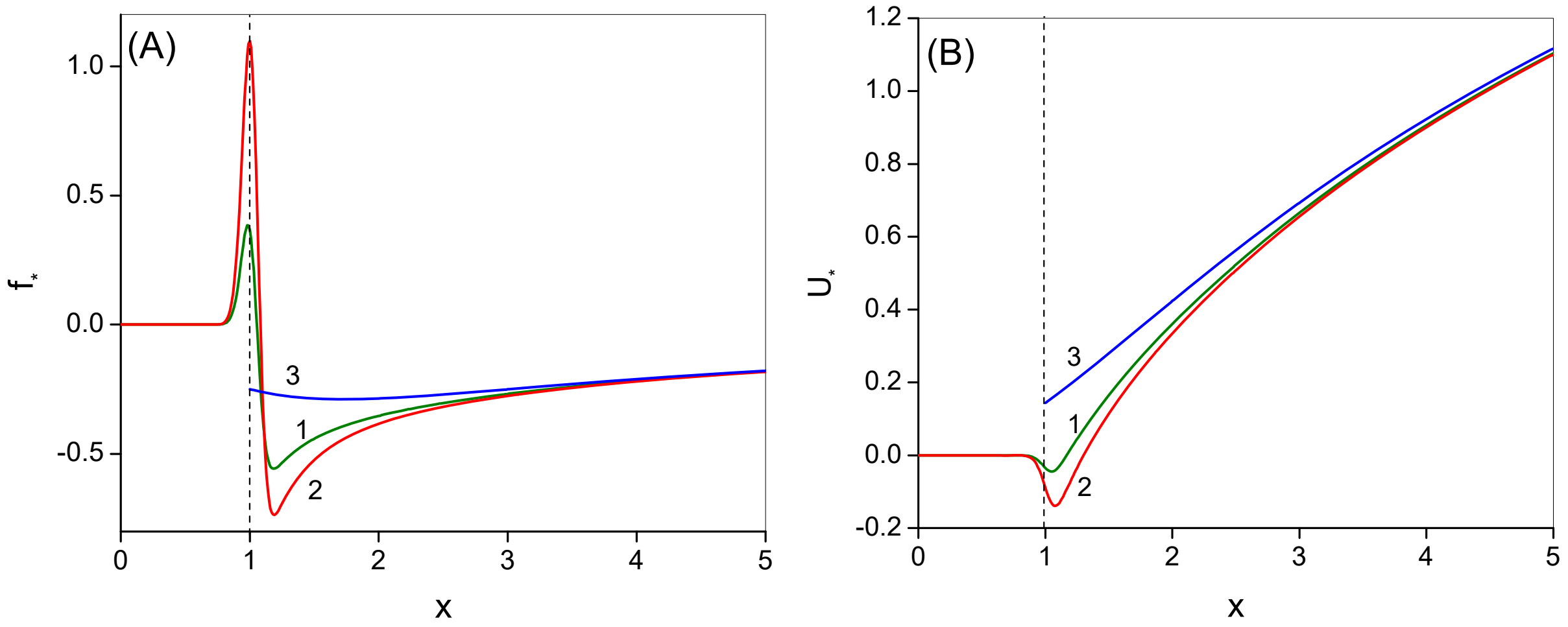

Fig. 3. (A) – $f_*$, (B) – $U_*$. Line 1 corresponds to $\eta_* = 5$, line 2 to $\eta_* = 10$, and line 3 to asymptotics (24). In the calculation: $\delta_* = 0.1$, $R_* = 1$. The vertical dashed lines in Fig. (A) and (B) separate the real vacuum from the fake one: to the left of them is the real vacuum; to the right is the false one.

Figure 4 shows the dependence of the depth of the potential well on $x$ and potentials for different values of $R_*$.

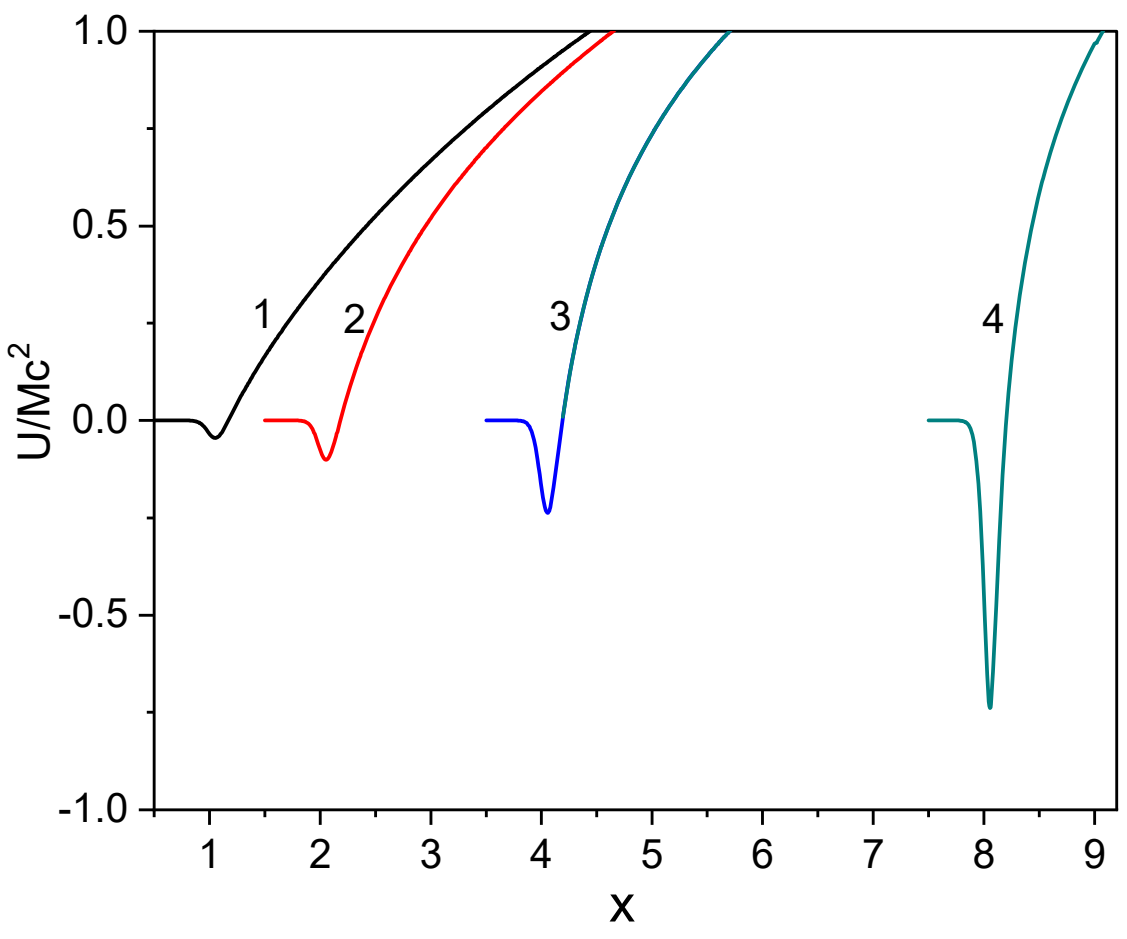

Fig. 4. Dependence of potential $U_*$ on $x$. Line 1 corresponds to $R_* = 1$, line 2 to $R_* = 2$, line 3 to $R_* = 4$, and line 4 to $R_* = 8$; $\eta_* = 5$. $\delta_* = 0.1$.

It should be noted that for all values of $x$, $g_{00} > 0$ only if $1 < \eta_* < \eta_{cr}$. If $\eta_* > \eta_{cr}$, the sign of $g_{00} = 1 - \frac{1}{r}\int_0^r r^2\left(\frac{8\pi G}{c^4}\tilde{T}_0^0 - \Lambda\right)dr$ in the transition region can change sign, which, in accordance with [13], indicates the presence of an event horizon. In this work, we will confine ourselves to only pointing out the possibility of the appearance of an event horizon in the transition region between the false and real vacuums. Figure 5 shows the dependence of the critical value $\eta_{cr}$ on $R_*$ .

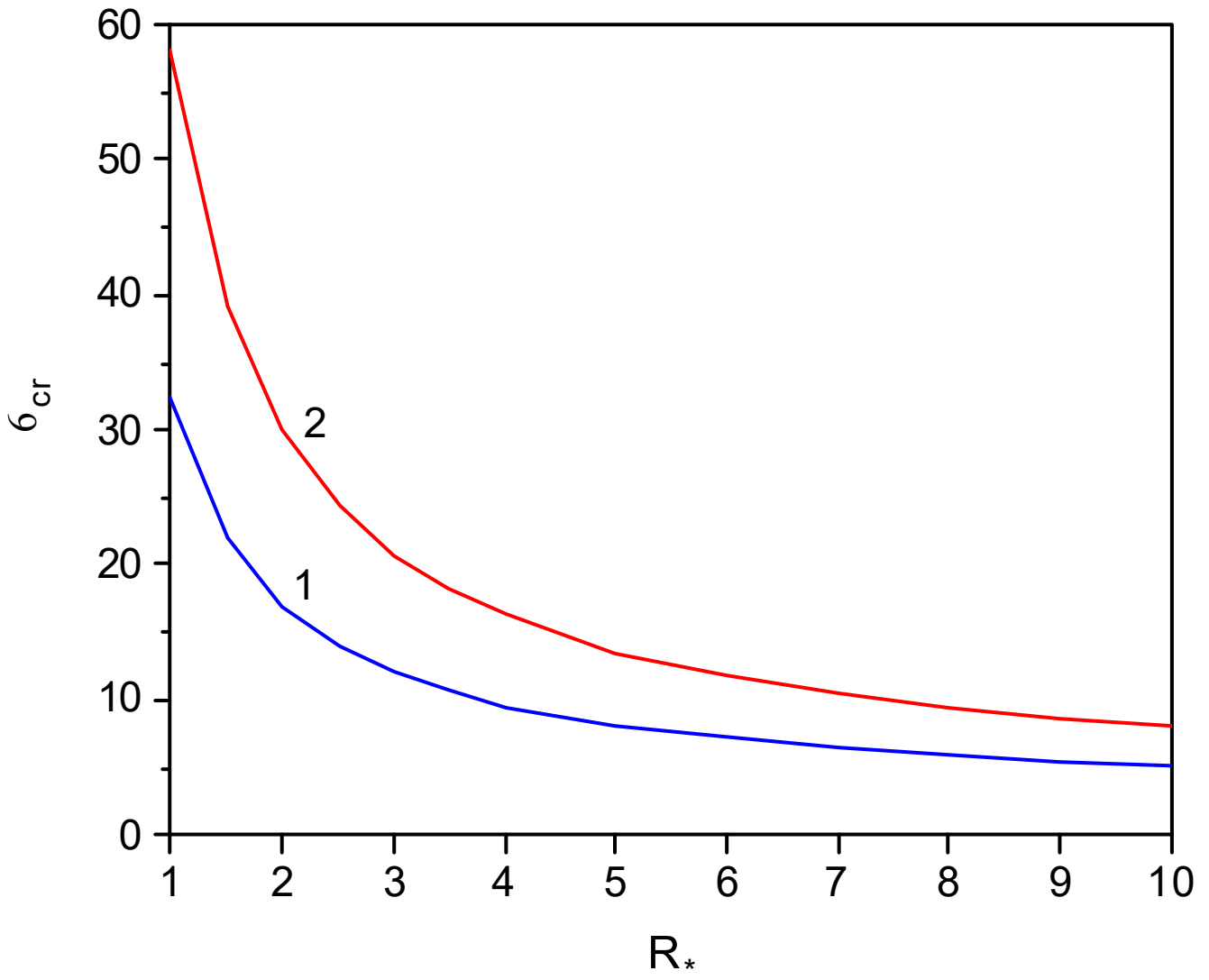

Fig. 5. Dependence of the critical value $\eta_{cr}$ on $R_*$. Line 1 corresponds to $\delta_* = 0.1$ and line 2 to $\delta_* = 0.05$.

## III. Formation of a thin layer of matter on the boundary of the physical and false vacuums

The transition layer separating the false vacuum from the real one (Fig. 1) can be considered as a mixture of a false vacuum, characterized by the value of $\Lambda$ and matter, characterized by the tensor $\tilde{T}_k^i$. This mixture, as shown above, forms a potential well for particles produced at the boundary, as shown in Figures 3B and 4B.

Since the issues of converting the energy of a false vacuum into the energy of generated matter and the dynamics of expansion of a bubble of a physical vacuum are beyond the scope of this article, we will assume below that a bubble of the physical vacuum expands at a constant rate.

It is known that massive and massless particles can be produced in a system with a significant gradient of the gravitational field (see, for example, [14-16]). Therefore, at the interface between the false vacuum and real vacuum the generation of particles can occur. The energy of the generated particles is drawn from the gravitational field and the energy of the false vacuum. In the case of Hawking evaporation of black holes, one particle is born beyond the event horizon and leaves the black hole while the other particle is born within the horizon and remains in the black hole. In our case of particle generation between the false and real vacuums,, two options are possible. If the energy of each of the generated particles is insufficient to overcome the potential barrier, then the particles remain inside the potential well in the transition region.

On the other hand, if the energy of the generated particles exceeds the potential barrier, then they leave the transition region. Note that the total momentum of the false vacuum and generated particles remains equal to zero. The possible cases described above are shown in Figure 6.

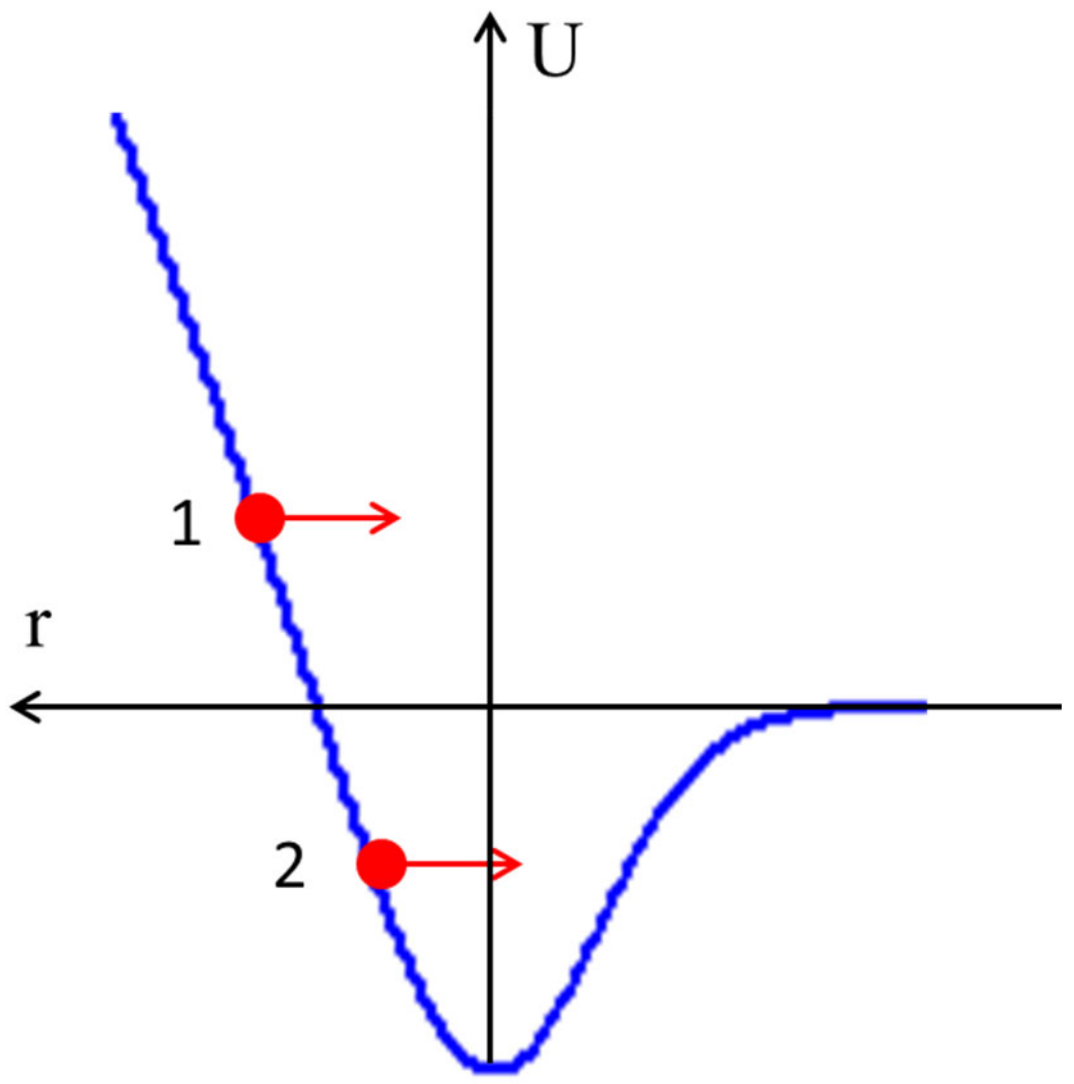

Fig. 6. Potential distribution on the boundary between the false and real vacuums. Particle 1, born at an interface, leaves the transition region and moves to the center of the bubble; particle 2, born inside the potential well, is captured and does not leave the transition region.

Consider the fate of the emitted particle. Potential wells at the boundaries of an expanding physical vacuum bubble are shown schematically in Figure 7. If the velocity of the expansion of the bubble boundary $u$ is less than the velocity of the mass particle $v$, then upon reaching the opposite wall of the bubble, the velocity of the particle in the frame of reference of the opposite wall will be less than that which it had at birth, and, accordingly, after elastic reflection from the wall, the kinetic energy of the reflected particle will decrease. These reflections from the walls will continue until the particle is captured either in the left potential well or the right one, shown in Figure 7, due, for example, to scattering on the captured particles.

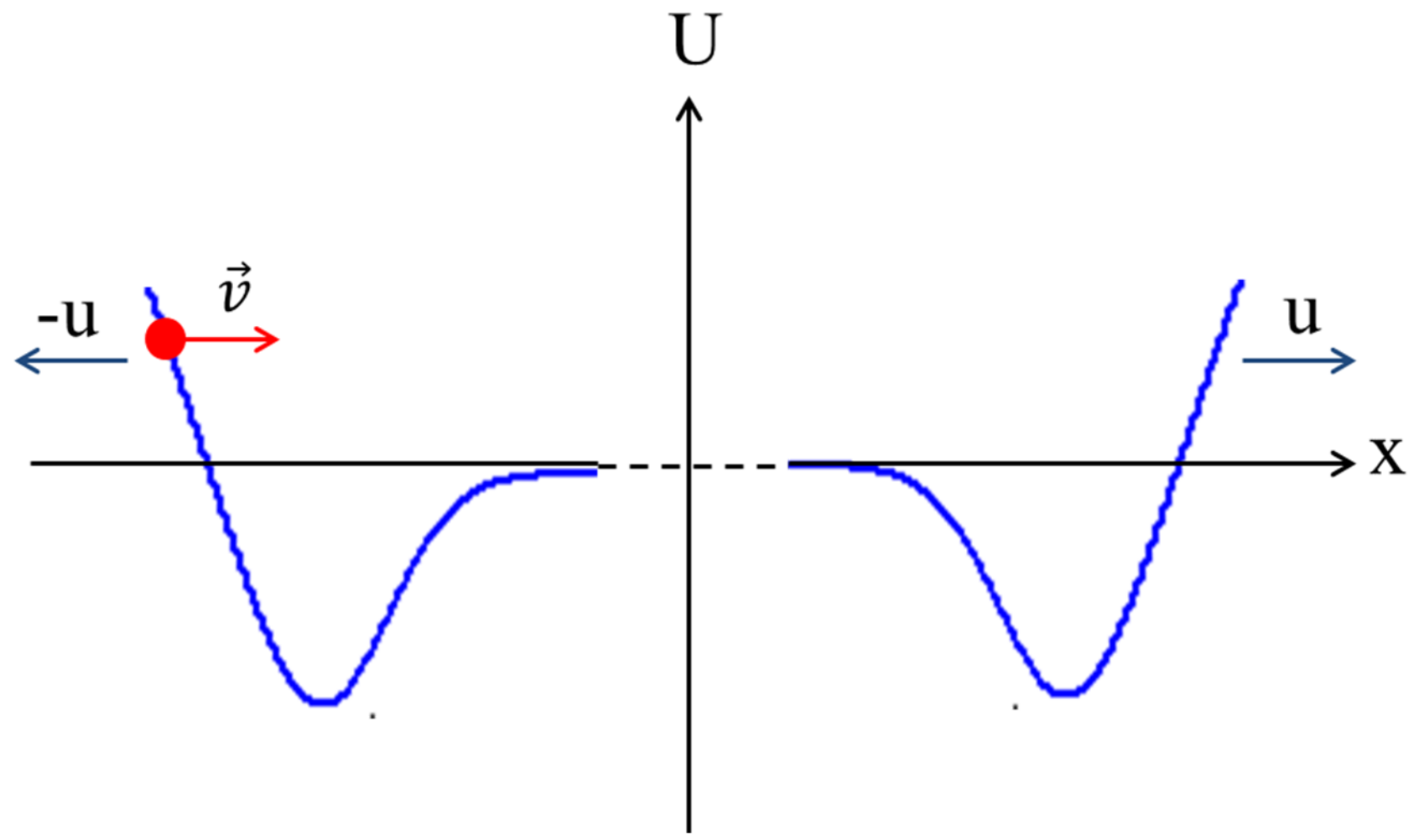

Fig. 7. Transitional layers of a physical vacuum bubble in a false one along the axis $x$ and the motion of a particle in a bubble of a physical vacuum expanding at a constant speed . After each elastic rebound, the particle velocity decreases by an amount equal to $2u$.

Let us estimate the rate of energy loss for a relativistic particle periodically colliding with the walls of a physical vacuum bubble expanding at a constant speed.

Without loss of generality, let us consider the simplest case, when the speed of the expansion of a bubble of a physical vacuum $u$ is constant and much less than the speed of light. In this case, collisions of particles with expanding bubble walls (Fig. 6) can be considered in the classical approximation. For simplicity, we will also assume that collisions of particles with walls are elastic. Let assume that in the transition layer on the left boundary, a particle was born, the speed of which is equal to $v$. Accordingly, its velocity in the laboratory frame of reference (associated with the center of the bubble) is equal to $v - u$ and associated with the right wall is equal to $v - 2u$ (the axis is directed from left to right). After an elastic collision with the right wall, the particle is reflected toward the left. Its speed in the frame of reference associated with the left wall is $-v + 4u$. It is obvious that after each elastic collision with the wall, the particle velocity in absolute value decreases by $2u$ in the frame of reference of the opposite wall. The particle deceleration process continues until the value of its velocity in the laboratory frame of reference becomes less than $u$.

The recurrent formulas connecting the time between collisions with the walls with the particle and the bubble expansion velocities after the $n$-th collision has the form

$$\begin{cases} v_{*n} = v_{*n-1} - 2u_* \\ 2R_{*n-1} + 2u_* t_{*n} = v_{*n} t_{*n} \end{cases}, \tag{25}$$

where $u_* = u/c$ is the bubble expansion rate, $R_{*n} = R_n/r_0$ is the bubble radius, $v_{*n} = v_n/c$ is the particle velocity after the $n$-th collision, $t_{*n} = t_n c/r_0$ is the time between collisions of a particle with the walls of the bubble, $R_{*0}$ is the bubble radius at the moment of particle birth, and $v_{*0}$ is the initial velocity of the particle in the frame of reference where it was born.

Table 1 shows an example of calculation by recurrent formulas (25) for the initial particle velocity $v_{*0} = 0.9$, the initial radius $R_{*0} = 1$, and $u_* = 0.05$.

**Table 1.**

| $n$ | $v_{*n}$ | $R_{*n}$ | $t_{*n}$ |
|---|---|---|---|
| 0 | 0.9 | 1 | 0 |
| 1 | 0.7 | 1.14 | 2.86 |
| 2 | 0.5 | 1.37 | 4.57 |
| 3 | 0.3 | 1.86 | 9.14 |
| 4 | 0.1 | 3.65 | 36.57 |

It is clear that taking into account the collision of particles with each other leads to the Maxwellization of the trapped particles (the appearance of temperature) and the filling of the potential well with trapped transit particles.

It should be noted that at the same values of $\eta_*$ and $\delta_*$ over time, the depth of the potential well (see Fig. 5) grows so that the energy distribution of particles in the potential well changes with time as $e^{-U(t)/k_B T}$, where $k_B$ is the Boltzmann constant, and $T$ is the temperature.

Obviously, this consideration takes place only for the case when $\eta_* < \eta_{cr}$. The emergence of an event horizon requires separate consideration. Note that, in contrast to a black hole, where $g_{00} = 1 - \frac{r_g}{r}$ has a singularity at zero, in our case there is no singularity. Figure 8 shows the dependence for $g_{00} =$

$\left(1 - \frac{1}{x}\int_0^x y^2(T^0_{*0} - \Lambda_*)dy\right)$ for $\eta_* < \eta_{cr}$ and $\eta_* > \eta_{cr}$. The points of intersection of $g_{00} = 0$ with curve 2 correspond to event horizons. As in the case of an ordinary black hole, $U_*$ and $f_*$ have discontinuities.

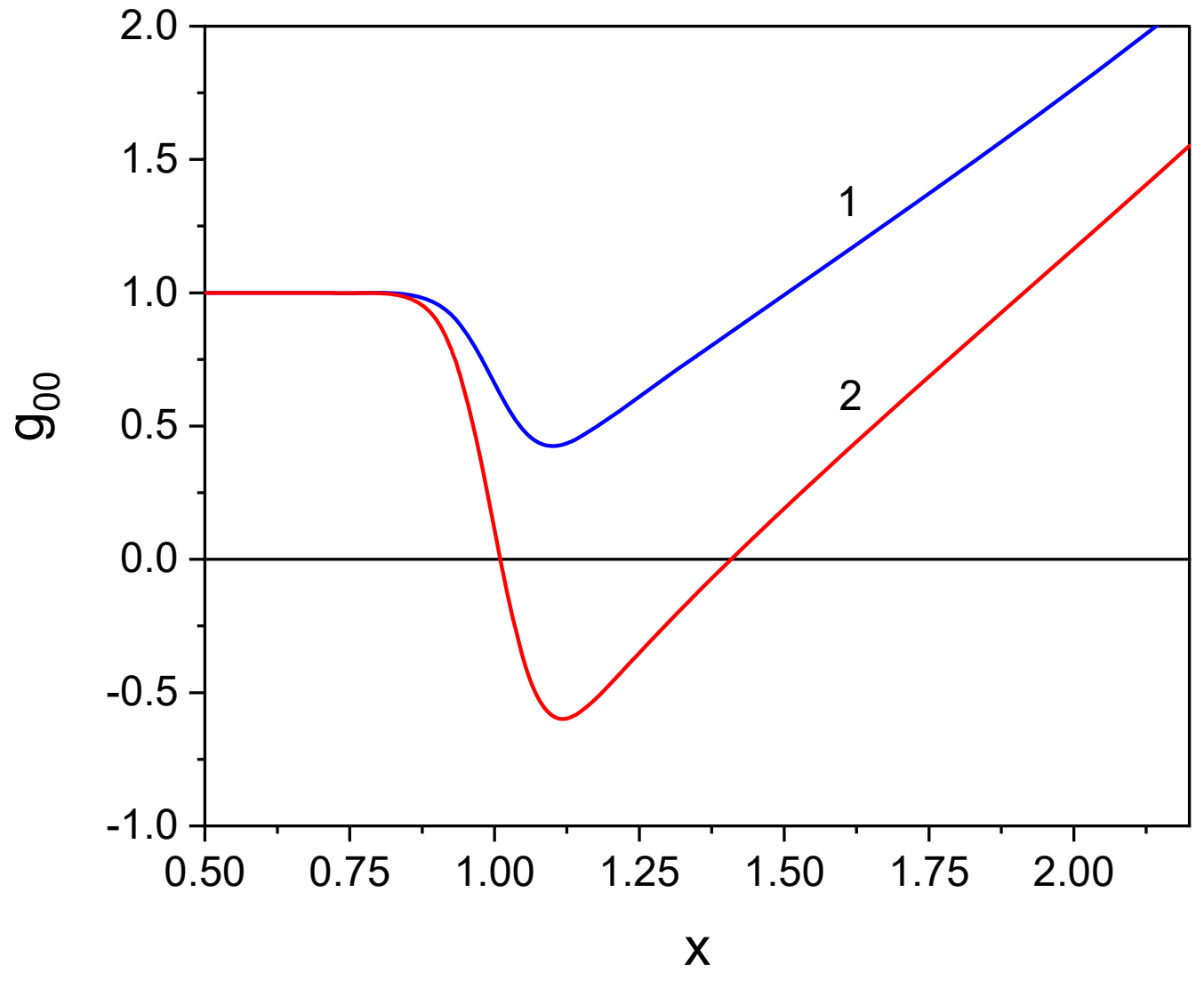

Fig. 8. Dependence $g_{00} = \left(1 - \frac{1}{x}\int_0^x y^2(T^0_{*0} - \Lambda_*)dy\right)$ on $x$. Line 1 corresponds to $\eta_* = 20 < \eta_{cr}$, $2 - \eta_* = 40 > \eta_{cr}$; $\delta_* = 0.1$. The points of intersection of $g_{00}$ with curve 2 correspond to event horizons.

## IV. Conclusions

The paper considers the transition layer at the boundary of the physical vacuum bubble, which arises in the false vacuum in the process of inflationary expansion. It is shown that in the transition layer, a potential well is formed for mass and massless particles that are generated in the region of an inhomogeneous gravitational field in the interface region. The mechanism of the formation of narrow layers filled with matter at the boundaries of the false and real vacuums is considered. These layers can serve as precursors to bridges in the observed large-scale cellular structure of the universe.

## References

[1] M.N. Shneider, M. Pekker, Cavitation model of the inflationary stage of Big Bang, Phys. Fluids, 33, 017116 (2021)

[2] A.H. Guth, The inflationary universe: a possible solution to the horizon and flatness problems, Phys. Rev. D 23, 347 (1981)

[3] M.N. Shneider, M. Pekker, "Liquid Dielectrics in an Inhomogeneous Pulsed Electric Field, Dynamics, Cavitation, and Related Phenomena (Second Edition)". (IOP Publishing, Temple Circus, Temple Way, Bristol, BS1 6HG, UK, 2019)

[4] A. D. Linde, Nonsingular regenerating inflationary Universe, Print-82-0554 (CAMBRIDGE). Full text can be found at http://www.stanford.edu/~alinde/1982.pdf

[5] A. D. Linde, A new inflationary universe scenario: a possible solution of the horizon, flatness, homogeneity, isotropy and primordial monopole problems, Phys. Lett. B 108, 389 (1982)

[6] A. D. Linde, Inflation can break symmetry in susy, Phys. Lett. B 131, 330 (1983)

[7] A. D. Linde, Chaotic inflation, Phys. Lett. B 129, 177 (1983)

[8] A. D. Linde, Eternally existing self-reproducing chaotic inflationary universe, Phys. Lett. B 175, 395 (1986).
[9] A. Vilenkin, The birth of inflationary universes, Phys. Rev. D 27, 2848 (1983)
[10] A. Vilenkin, Predictions from quantum cosmology, Phys. Rev. Lett. 74, 846 (1995)
[11] A. Vilenkin, Making predictions in eternally inflating universe, Phys. Rev. D 52, 3365 (1995)
[12] A. Vilenkin, "Many Worlds in One: The Search for Other Universes", (New York: Hill and Wang, 2007)
[13] L.D. Landau, E.M. Lifshitz, "The Classical Theory of Fields", 4th Edition, Course of Theoretical Physics, (Butterworth-Heinemann (Elsevier), 1975)
[14] L. Parker, Quantized fields and particle creation in expanding universes. I, Phys. Rev., 183, 1057 (1969)
[15] Ya.B. Zel'dovich, L.P. Pitaevskii, On the possibility of the creation of particles by a classical gravitational field, Comm. Math. Phys, 23, 185 (1971)
[16] Ya.B. Zel'dovich, A.A. Starobinskii, Particle production and vacuum polarization in an anisotropic gravitational field, Soviet Physics JETP, 34, 1159 (1972)